\documentclass{article}

\usepackage{arxiv}

\usepackage[utf8]{inputenc} 
\usepackage[T1]{fontenc}    
\usepackage{hyperref}       
\usepackage{url}            
\usepackage{booktabs}       
\usepackage{amsfonts}       
\usepackage{nicefrac}       
\usepackage{microtype}      
\usepackage{lipsum}		
\usepackage[monochrome]{xcolor}
\usepackage{graphicx}
\usepackage[authoryear]{natbib}
\usepackage{doi}
\usepackage{setspace} 
\usepackage[hashEnumerators,smartEllipses]{markdown}

\title{\textcolor{blue}{Data-centric Engineering: integrating simulation, machine learning and statistics. Challenges and Opportunities}}

\author{ \href{https://orcid.org/0000-0002-9624-5146}{\includegraphics[scale=0.06]{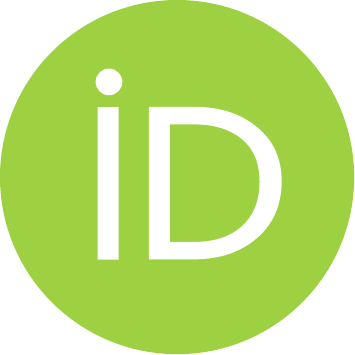}\hspace{1mm}Indranil Pan} \\
	Imperial College London, UK, SW7 2AZ\\
	The Alan Turing Institute, UK, NW1 2DB\\
	\texttt{i.pan11@imperial.ac.uk} \\
	\And
	\href{https://orcid.org/0000-0002-7644-931X}{\includegraphics[scale=0.06]{orcid.pdf}\hspace{1mm}Lachlan~R.~Mason} \\

	The Alan Turing Institute, UK, NW1 2DB\\
	Imperial College London, UK, SW7 2AZ\\
	\texttt{l.mason@imperial.ac.uk} \\
	\And
	\href{https://orcid.org/0000-0002-0530-8317}{\includegraphics[scale=0.06]{orcid.pdf}\hspace{1mm}Omar~K.~Matar} \\
	Imperial College London, UK, SW7 2AZ\\
	The Alan Turing Institute, UK, NW1 2DB\\
	\texttt{o.matar@imperial.ac.uk} \\

}

\renewcommand{\shorttitle}{\textit{arXiv} Template}

\hypersetup{
pdftitle={Data-centric Engineering: integrating simulation, machine learning and statistics. Challenges and Opportunities},
pdfauthor={Indranil Pan, Lachlan Mason, Omar Matar},
pdfkeywords={Digital twins, Artificial Intelligence, CFD, FEM},
}

\begin{document}
\maketitle

\doublespacing
\begin{abstract}
\textcolor{blue}{Recent advances in machine learning, coupled with low-cost computation, availability of cheap streaming sensors, data storage and cloud technologies, has led to widespread multi-disciplinary research activity with significant interest and investment from commercial stakeholders. Mechanistic models, based on physical equations, and purely data-driven statistical approaches represent two ends of the modelling spectrum. New hybrid, data-centric engineering approaches, leveraging the best of both worlds and integrating both simulations and data, are emerging as a powerful tool with a transformative impact on the physical disciplines. We review the key research trends and application scenarios in the emerging field of integrating simulations, machine learning, and statistics. We highlight the opportunities that such an integrated vision can unlock and outline the key challenges holding back its realisation. We also discuss the bottlenecks in the translational aspects of the field and the long-term upskilling requirements for the existing workforce and future university graduates.}
\end{abstract}


\vspace{3mm}
\section{Introduction}
\vspace{3mm}
Recent advances in the field\textcolor{blue}{s} of machine learning (ML) and artificial intelligence (AI) in the last decade \textcolor{blue}{have} elicited interest from diverse groups including the scientific community, industry stakeholders, governments and society at large \citep{royal2017machine}. There has been a frenzy of investment and startup activity to capitalise on these advances 
\citep{Forbes2020}.
Historically, however, the field of AI has gone through multiple peaks of inflated expectations and consequent disillusionment – in the 1970s and 1980s, for example – resulting in the `AI winter' which saw massive funding cuts, limited adoption by industries, and \textcolor{blue}{the} end of focused research activity in the area \citep{AIwinter95}.
There is an ongoing debate about whether \textcolor{blue}{the recent interest in AI} is yet another over-hyped phenomenon which will fizzle out soon. Indeed, Gartner hype-cycle reports \citep{Gartner2019}  \textcolor{blue}{point to evidence that AI may be going through a peak of inflated expectations by tracking} multiple key-words associated with the current boom (e.g.~deep learning, chatbots, machine learning, \textcolor{blue}{and} AutoML). Researchers in \textcolor{blue}{Big Tech companies, however}, believe that unlike other times, AI has already added a lot of value, become central to their product strategies, and is here to stay \citep{chollet2018deep}. The academic community \textcolor{blue}{across} engineering disciplines, where there is potential for uptake of recent advances in AI \textcolor{blue}{(e.g.~aeronautical, chemical, and mechanical)}, take a more nuanced stand on the potential of the current AI hype.
While there is acknowledgement that adoption of purely data-driven ML approaches can address a number of challenges, \textcolor{blue}{researchers} \textcolor{blue}{support the view} that the key to transform\textcolor{blue}{ing} these disciplines \textcolor{blue}{involves a {\it data-centric engineering} approach; this involves} exploiting domain\textcolor{blue}{-}specific knowledge and integrating mechanistic models\textcolor{blue}{,} or other forms of symbolic reasoning, 
with data-driven processing \citep{venkatasubramanian2019promise}. Additionally, there are multiple concerns regarding the black-box nature of  deep-learning algorithms, poor integration with prior knowledge, \textcolor{blue}{and} trustworthiness of the solutions among others \citep{marcus2018deep}.

Deep learning, which has spurred the current resurrection of AI, has already beaten multiple benchmarks in image and speech recognition, drug design, analysis of particle accelerator data, genetics and neuroscience \citep{lecun2015deep}. As pointed out in an extensive review on historical developments in deep learning \citep{schmidhuber2015deep}, current deep-learning technology components have existed for more than three decades now, including multiple successive non-linear layers, back propagation algorithms and convolution neural networks. Although there have been algorithmic improvements, the key reason why deep learning is beating a lot of benchmarks is because current graphics processing unit (GPU) assisted computers have millions of times more computing power than desktops of 1990s \citep{schmidhuber2015deep}. Coupled with \textcolor{blue}{a rise} in computing power and neural network sizes, the increase in availability of large scale labelled \textcolor{blue}{data} is another key reason for success of deep-learning algorithms \citep{sun2017revisiting}.  
However, deep learning in its current form is data and compute hungry and recent estimates indicate that further improvements in performance of such systems are becoming economically, technically, and environmentally unsustainable \citep{thompson2020computational}. Given that improvements in hardware performance are slowing, the authors \citep{thompson2020computational} project that progress will depend on more computationally efficient methods, to which either deep learning or newer algorithmic learning methods will have to adapt.

The initial success of deep learning in image recognition, speech and text processing has \textcolor{blue}{attracted} the attention of researchers in traditional engineering fields. Multiple review papers consolidating trends in individual engineering disciplines have been published\textcolor{blue}{; t}hese include applications of AI/ML in chemical process systems engineering \citep{lee2018machine}, fluid mechanics \citep{brunton2020machine}, smart energy systems \citep{lund2017smart}, smart cities \citep{o2019smart}, structural health monitoring \citep{flah2020machine} in civil engineering applications, engineering risk assessment \citep{hegde2020applications}, process systems safety \citep{goel2020integration}, bio-chemical industries \citep{udugama2020role}, bio-energy systems \citep{liao2021applications}, industrial monitoring and process control \citep{gopaluni2020modern} among others. 

There has been some initial success in applying ML/AI models in a plug\textcolor{blue}{-}and\textcolor{blue}{-}play fashion without  \textcolor{blue}{the requirement of modifying} the underlying algorithms for domain-specific applications. However, targeting the more challenging problems in each discipline requires customising the ML/AI algorithms to incorporate domain knowledge alongside data\textcolor{blue}{-}driven methods \citep{venkatasubramanian2019promise}. In engineering domains, especially in the prototyping design phase, very little data \textcolor{blue}{are} typically available since there is no operational plant or system to generate data in the first place. Physics-driven models are much more useful in such situations compared with deep-learning methods, which are data hungry.
Moreover, purely data-driven approaches do not encode physical laws such as conservation of mass, momentum or energy, which form the fundamental basis of any engineering application. Therefore, operations engineers are skeptical of decision-making based on outputs of such models. Deep-learning models, for example, are highly vulnerable to adversarial examples, which are almost imperceptible to humans, but can easily fool the ML model and cause it to misclassify \citep{kurakin2016adversarial}. Such erroneous outputs can have catastrophic consequences in a safety critical engineering environments with long term financial and legal implications. \textcolor{blue}{In fact, an active area of research \citep{zhang2019adversarial, yuan2019adversarial} in the deep-learning community involves} generating adversarial examples and \textcolor{blue}{providing} defences against them. 

Alongside \textcolor{blue}{the} trust issues \textcolor{blue}{mentioned}, there are also \textcolor{blue}{other issues} related to interpretability of such data-driven ML models.
In general, there is a tension between accuracy and interpretability of models. Physics-driven simulations, based on bottom-up modelling, are useful for interpretability via the insights they provide. Although the simulation predictions may not provide a perfect match to physical phenomena due to the use of simplifying assumptions, they do, nonetheless, capture the overall observed trends. Data-driven approaches, on the other hand, give a very good predictive performance within the regime of the training dataset, though often fail drastically to generalise to input regimes outside of this dataset.
Furthermore, unlike \textcolor{blue}{models generated from the equations governing underlying physical laws}, most ML models are black boxes \textcolor{blue}{whose use} hinders intuitive understanding. 
\textcolor{blue}{Moreover, it is unclear} how such models will behave for sets of inputs \textcolor{blue}{outside of the training dataset}. 
Current research, therefore, focusses on interpretability of trained ML models \citep{molnar2020interpretable}.
%
%
Techniques such as local interpretable model agnostic explanations (LIME) \citep{ribeiro2016should} and Shapely additive explanations (SHAP) \citep{lundberg2017unified} support a post-hoc analysis mapping feature importance for particular prediction outcomes. However, the notion of interpretability is itself ill-defined and post-hoc interpretations of models can be potentially misleading \citep{lipton2018mythos}.

\begin{figure}[h!]
  \centering
  \includegraphics[width=14cm]{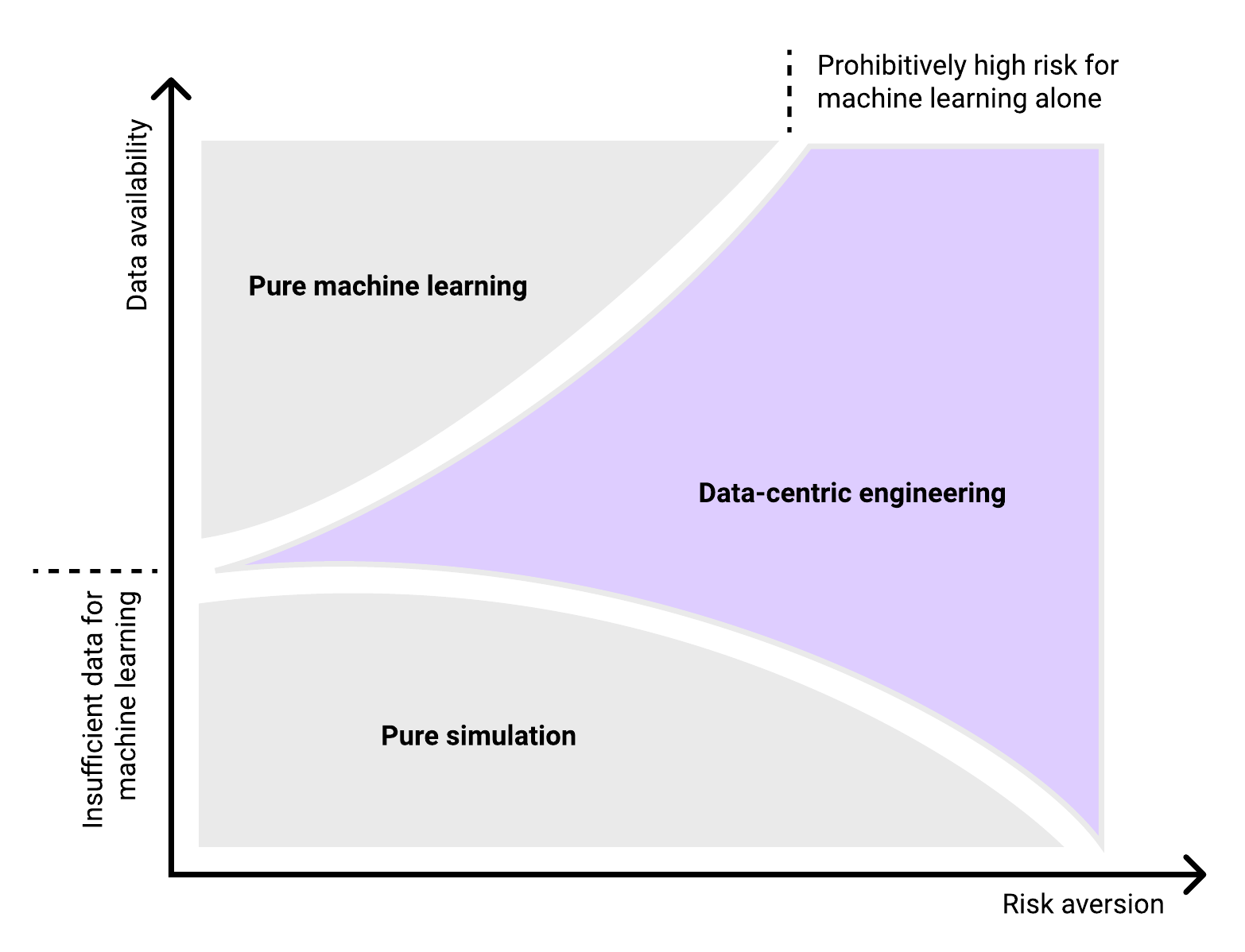}
  \caption{\textcolor{blue}{Scope for data-centric engineering, where data-driven and physics based models are integrated, in terms of risk aversion and data availability. Regimes appropriate for pure machine learning and pure simulation are shown.}}
  \label{fig:quad}
\end{figure}

\textcolor{blue}{Problems that require data-centric engineering solutions can be mapped in the space of data availability and risk aversion, as shown in Figure \ref{fig:quad}. Typically, pure physics-driven simulations are employed in cases where there is insufficient data for ML and the governing physical laws are well understood, as shown by the `pure simulation' region in the lower half of the graph. Typical ML approaches are for cases where data availability is high and risks associated with failed predictions are low, as shown by the `pure machine learning' region in the upper-left region. Any application which has material financial, legal or other hazardous downsides generally needs integration of both data and simulations to de-risk decisions. Problems lying in these zones (indicated by the `data-centric engineering' region to the right) benefit the most from data-centric engineering approaches.}

The integration of physics-based models with data-driven \textcolor{blue}{ML approaches} through data-centric engineering offer\textcolor{blue}{s} a good \textcolor{blue}{trade-off} in terms of both interpretability and fit to real world data. Such integrations are also more data-efficient as opposed to pure deep-learning techniques, for example. While more classical approaches, such as statistical calibration of model parameters of physics-driven models, have been around for a long time, emerging approaches \textcolor{blue}{are gaining traction; these include} physics-informed neural networks \citep{raissi2019physics}, encoding physics laws in the \textcolor{blue}{ML} model itself. In the following sections, we \textcolor{blue}{highlight emerging trends in the field and outline }generic use cases \textcolor{blue}{featuring} integration \textcolor{blue}{of} data, physics simulations, statistics, and ML.\textcolor{blue}{ Relevant papers for chemical engineering applications are also cited for each of these use cases.} \textcolor{blue}{The objective of this paper is hence to provide a high-level overview without going into in-depth implementation details for each of the techniques. 
} 
\section{Integration of simulation, machine learning and statistics: current approaches}
\label{sec:integration_approaches}

Classically\textcolor{blue}{,} numerous methodologies have been developed in the fields of both frequentist and Bayesian statistics to handle model parameter estimation, calibration, design of experiments, and model comparison. However, most of the underlying models used in traditional statistics are computationally inexpensive as compared with spatio—temporally resolved engineering models (e.g.~\textcolor{blue}{those based on the use of computational fluid dynamics }or \textcolor{blue}{finite-element} simulations). Moreover, legacy engineering codes might only be available through a function call (without access to modifications of the internals of the codes) and access to underlying gradient \textcolor{blue}{information} might not be readily available. Therefore\textcolor{blue}{,} practical integration of computationally expensive engineering simulations with classical statistical methodologies require\textcolor{blue}{s} adjustments to the \textcolor{blue}{statistical} algorithms to make fewer function calls and ensure computational tractability. The following sub-sections give a brief overview of such integrated application scenarios.       

\subsection{Surrogate modelling and active learning}

Training computationally inexpensive statistical or ML surrogates using data generated from complex simulation models is a common approach for ensuring computational tractability in engineering design \citep{forrester2008engineering}. Once trained, such surrogate models can be used to predict simulation outcomes which are not originally in the training dataset, or used in an optimisation setting. However, since the surrogates are typically data-driven models, the predictions outside the training regime can be error prone. The \textcolor{blue}{surrogates} also introduce artificial or false local minima and hence should be used with care in an optimisation framework \citep{jin2000evolutionary}. 

Data generation for training a surrogate model generally involves batch sampling of the expensive simulation model based on a pre-generated sampling scheme (e.g.~Latin Hypercube sampling or other space-filling designs). Active learning on the other hand can take sequential decisions on which point to sample based on prior samples and the corresponding simulator output. Such a scheme gives better surrogate model performance with fewer calls to the expensive simulator for training \citep{settles.tr09}. Similar ideas in the statistics literature \textcolor{blue}{are} known as `optimal experimental design'. 

\subsection{Calibration of simulation models}

Calibration refers to the process of adjusting the parameters of a simulation model to fit \textcolor{blue}{an} observed dataset. Once a simulation model is calibrated in this fashion, the estimated parameters are used in the simulation to obtain prediction\textcolor{blue}{s} in regimes where data \textcolor{blue}{are} \textcolor{blue}{un}available. Frequentist approaches rely on obtaining the maximum likelihood estimate of the parameters by directly evaluating the expensive simulation sequentially within an optimisation routine \citep{vecchia1987simultaneous}. Surrogate models can be used in such an optimisation routine to \textcolor{blue}{accelerate} the process, but they can introduce substantial uncertainty in estimation \citep{wong2014frequentist}. Bayesian calibration of expensive computer simulation models \citet{kennedy2001bayesian}, on the other hand, obtains full posterior distributions of the simulation model parameters instead of point estimates. Typically Gaussian Processes are used as surrogate models, although other efficient surrogate models can also be used. The Bayesian approach allows for incorporation of all sources of uncertainty and also attempts to correct for the inadequacy of the simulation model itself \citep{kennedy2001bayesian}.
\textcolor{blue}{A related but slightly different context in chemical engineering is the calibration of non-linear dynamic process models using system identification techniques mostly for process control \citep{ljung1983theory}. The models might be coupled ordinary differential equations or time series auto-regressive models and are generally not computationally expensive. Some early works on applying system identification for chemical processes include applications in paper machines \citep{astrom1967computer}, boilers \citep{eklund1969multivariable}, heat exchangers \citep{liu1987nonlinear}, integrated chemical plants \citep{garcia1981optimal}, among others. It is possible to learn entire calibration curves using neural networks instead of just a few model parameters as well. }

\subsection{Data assimilation}

Data assimilation is \textcolor{blue}{a form of recursive Bayesian estimation and has roots in Kalman Filtering \citep{kalman1960new} in control theory from the 1960s.} It originated in the field of numerical weather prediction to optimally combine the dynamical models of atmospheric systems with observation data to make forecasts. Data assimilation has been extensively studied and has a long history since the 1980s with multiple in-depth review papers on the subject \citep{ghil1991data, navon2009data, bannister2017review}. Discretised versions of spatio-temporal partial differential equations have a large number of variables which make computations involving high dimensional covariance matrices intractable. To overcome such issues, techniques such as ensemble Kalman filtering implement a Monte Carlo version of sequential Bayesian updating \textcolor{blue}{and} have been very popular \citep{evensen2003ensemble}. Data assimilation is \textcolor{blue}{clearly} not limited to weather prediction problems only and can be applied to any dynamical system simulator. 

\subsection{Simulation-assisted data generation for machine learning}

Synthetic data can be generated from virtual simulation models or \textcolor{blue}{other validated} physical models which can then be \textcolor{blue}{combined with operational data available from real world systems and then} used for training machine learning models \citep{klein2018data}. Such techniques are especially useful in scarce-data regimes, for example, failure modes in aircraft gas turbines \citep{saxena2008damage} that are financially very expensive if they \textcolor{blue}{occur} on a real system. 
\textcolor{blue}{This is similar to surrogate modelling but with the specific goal of trying to improve ML models by improving the range of the available dataset (hence enhancing the predictive accuracy of these models in these regimes) and overcoming issues of class imbalance (for classification problems, for example) for ML model training.}  Other cases of simulation-assisted ML involves scenarios where the \textcolor{blue}{underlying physics is not sufficiently well understood to obtain high fidelity simulations and only a small number of training data points are available} \citep{deist2019simulation}.


\subsection{Design of experiments}

Design of experiments (DoE) has a long history in statistics, starting \textcolor{blue}{with the work} of R.A.~Fisher \citep{fisher1937design} and was used for laboratory and field experiments. DoE devises strategies for conducting a \textcolor{blue}{set of experiments} which will yield maximum information (in a statistical sense) for parameter estimation and model validation \citep{franceschini2008model}. Adaptations of DoE sampling algorithms to deterministic numerical simulations and corresponding algorithmic packages have been reviewed in \citet{giunta2003overview}. Model-based DoE aims to use model equations and current parameters within an optimisation framework to predict the information content of the subsequent experiment \citep{franceschini2008model}. 

\subsection{Inverse problems}

In the context of numerical simulations, \textcolor{blue}{inverse problems refer to  finding the input variables \textcolor{blue}{(or parameters of the simulation model)} which can be used in the simulator to obtain the observed output.} There are both frequentist and Bayesian approaches to \textcolor{blue}{the solution of} inverse problems \citep{vogel2002computational}. \textcolor{blue}{Indeed, in the frequentist context which tries to obtain best possible parameters or input variables, inverse problems can be seen essentially as optimisation problems. For the Bayesian case, the answer is a distribution of each parameter instead of a single point estimate}. Generally, the solution of inverse problems requires multiple calls to the expensive forward numerical simulator, which can quickly become computationally intractable. Strategies such as using a surrogate model for the forward simulator, reducing the dimensions of the input space, or more efficient sampling techniques (e.g.~better Markov Chain Monte Carlo schemes in a Bayesian setting) are adopted in such situations \citep{frangos2010surrogate}. Solving inverse problems is difficult due to non\textcolor{blue}{-}existence or non-uniqueness of solutions or high sensitivity of the solutions to small changes in inputs.  

\subsection{Optimisation of engineering processes and design}
\vspace{2mm}
Optimisation of \textcolor{blue}{model parameters} to maximise a performance metric is at the heart of any engineering design or operational problem. Optimisation can be applied at multiple levels, e.g.~at the individual engineering component level or \textcolor{blue}{at the} overall system-level design. Efficient solution methods exist for convex optimisation problems \citep{boyd2004convex}. Such methods, however, are not applicable in simulation-based optimisation cases (i.e. where the objective function requires sampling a complex simulation model instead of being expressed as a set of algebraic equations). Complex simulation models might also be discontinuous, have multiple local minima \textcolor{blue}{and} no access to gradient information (due to \textcolor{blue}{the use of} legacy codes \textcolor{blue}{that} can only be queried through a function call), which further impedes \textcolor{blue}{the} use of traditional efficient optimisation methods. 
Surrogate model based optimisation is useful in such scenarios and is \textcolor{blue}{often termed} `simulation-based optimisation' or `meta model-assisted optimisation'. Efficient global optimisation \citep{jones1998efficient} using Kriging or Gaussian Process surrogate models has been widely used in this context. In the ML community, similar surrogate modelling methods have been used and \textcolor{blue}{are} commonly referred to as `Bayesian optimisation' \citep{shahriari2015taking}. Other strategies involve training multiple surrogate models and using different model management strategies to decide the best point to evaluate the expensive objective function \citep{goel2007ensemble} in each iteration. Surrogate-assisted evolutionary optimisation is another emerging field which is useful in solving computationally expensive single and multi-objective optimisation problems. These \textcolor{blue}{optimisation strategies} have also been promising for dynamic, constrained or multi-modal optimisation problems \citep{jin2011surrogate}. 

\subsection{Sensitivity analysis and forward uncertainty propagation for simulation models}

Sensitivity analysis aims to attribute the uncertainty in a simulation model to the different sources of uncertainty in the model inputs \citep{saltelli2004sensitivity}. An exposition of experimental designs and algorithms for sensitivity analysis is detailed in \citet{saltelli2008global}.  
A related concept is the forward propagation of uncertainty from inputs of \textcolor{blue}{a} dynamical system to its outputs. Random Monte Carlo sampling is one approach to solve the problem. Polynomial chaos expansions (PCE) and related methods \citep{xiu2010numerical} have been shown to work well in low-dimensional cases.



\section{Emerging research areas \textcolor{blue}{in data-centric engineering}}

Taking advantage of recent algorithmic advances and widely available computing power, tighter integrations are emerging between simulations, statistics, and machine learning \textcolor{blue}{ with a data-centric engineering approach.} We highlight the following emerging research themes in this area which are gaining traction. 

\subsection{Digital twins: Old wine in a new bottle?}

\textcolor{blue}{Though multiple definitions exist, a digital twin can be contextualised as
\begin{quote}
a set of virtual information constructs that mimics the structure, context and behavior of an individual or unique physical asset, that is dynamically updated with data from its physical twin throughout its life-cycle, and that ultimately informs decisions that realize value. 
\newline 
— \citet{aiaa2020digital}
\end{quote}
Digital twins are not only deployed in engineering settings but also in diverse fields including healthcare and information systems \cite{niederer2021scaling}.} \textcolor{blue}{According to} \citet{wright2020tell}, the digital-twin concept is an amalgamation of several existing mature concepts. 
Critics argue that having streaming data from a physical asset and updating simulation models for monitoring and control has already existed for decades in \textcolor{blue}{the} engineering industries. The novelty is related to the confluence of higher computational power, cheap sensors, and cloud technologies, which can run more powerful models and algorithms, and process larger amount\textcolor{blue}{s} of data leading to much richer insights and intervention policies. Such a confluence also paves the way to a connected ecosystem of digital twins as opposed to individual twins for a particular unit or process operation. Countries such as the UK have adopted national digital-twin programme\textcolor{blue}{s} for such connected digital twins to deliver value to society, the economy, and the environment \citep{nationalDT2018}.  

\subsection{Hybrid models using machine learning and dynamical systems}

Hybrid paradigms \textcolor{blue}{that}  integrate \textcolor{blue}{ML and simulators}, \textcolor{blue}{which are based on the solution of dynamical systems comprising ordinary and/or partial differential equations (PDEs),} are emerging as a powerful tool. \textcolor{blue}{Historically in chemical engineering, there have been multiple attempts over decades to couple neural networks with first-principles process models. \citet{psichogios1992hybrid} used a hybrid neural network to model a fed-batch bioreactor, which is more interpretable than standard neural networks, can interpolate and extrapolate more accurately, and require fewer training samples. \citet{kramer1991nonlinear} introduced non-linear principal component analysis (now more popularly known as auto-encoders in the ML community) to obtain lower-dimensional feature representations of the underlying dynamical system and showed its successful application with time-dependent batch reaction data obtained from first-principles reaction engineering simulations. 
\citet{rico1994continuous} introduced gray-box identification for partially known first-principles models of nonlinear dynamical systems using neural networks and applied it to a model reacting system. 
Early successful attempts at solving ordinary differential equations (ODEs) and PDEs using neural networks was proposed in \citet{lagaris1998artificial}. The method was shown to have superior performance as compared to standard finite-element methods and can scale to high-dimensional problems.
Identification of distributed parameter systems using neural networks and PDEs from spatio-temporally resolved sensor data was proposed in \citet{gonzalez1998identification}. The method can exploit partial knowledge of the underlying PDE and is better suited than lumped parameter models, which can be under-resolved and miss salient features of the underlying system \citep{gonzalez1998identification}. 
\citet{rico1992discrete} introduced neural network schemes for identifying long-term time series predictions for commonly observed phenomena in ODEs, including  bifurcation and temporally complicated periodic behaviour. They validated the methodology with experimental data for electro-dissolution.}

\textcolor{blue}{More recently,} physics-informed neural networks (PINNs) \citep{raissi2019physics} provide a solution scheme for training deep neural networks while respecting physical laws expressed by nonlinear \textcolor{blue}{PDEs}. Such schemes are data-efficient due to incorporation of physical laws and \textcolor{blue}{have} been extended to incorporate other cases such as solving fractional PDEs \citep{pang2019fpinns}, variational solutions of PINNs \citep{kharazmi2019variational, khodayi2020varnet}, physics-informed generative adversarial networks \textcolor{blue}{(GANs)} for solving stochastic differential equations \citep{yang2020physics} among others. \citet{berg2018unified} approximated solutions to PDEs in complex geometries using deep learning where classical mesh-based methods cannot be used \textcolor{blue}{due to complicated polygons and short line segments in the geometry which places severe restrictions on domain discretisation by triangulation}. 

Neural \textcolor{blue}{ODEs} are a new family of deep neural networks employing standard ODE solvers as a model component within the network \citep{chen2018neural}. The scheme is memory efficient and can explicitly control the trade-off between numerical accuracy and computational speed. \citep{chen2018neural}. Improvisations and extensions of the scheme include graph neural ODEs \citep{poli2019graph} \textcolor{blue}{and} Bayesian versions \citep{dandekar2020bayesian} among others. 

\subsection{Probabilistic numerics}

Probabilistic numerics is an emerging field where numerical tasks such as integration, optimisation, and solutions of differential equations can report uncertainties (e.g.~arising from loss of precision due to hardware constraints or other approximations) along with their solutions \citep{hennig2015probabilistic}. Numerical tasks interpreted as an inference problem in this way might pave the way for propagating both computational errors and inherent uncertainty across chains of numerical methods applied sequentially, helping to monitor and actively control computational effort \citep{hennig2015probabilistic}.

\subsection{Probabilistic programming for simulations}

Probabilistic programming languages (PPLs) allow users to specify statistical models (along with observations) and 
to perform inference with minimal user intervention. PPLs including WinBugs \citep{lunn2000winbugs}, Stan \citep{carpenter2017stan} and PyMC3 \citep{salvatier2016probabilistic} have popularised Bayesian statistical inference to a wider audience. The underlying probabilistic models commonly used in the statistical literature \textcolor{blue}{are} generalised linear models, hierarchical models \textcolor{blue}{and non-parametric models}. Increasingly, PPLs are being coupled to simulation models to \textcolor{blue}{undertake} simulation-based inference, which might have a long lasting impact on science \citep{cranmer2020frontier}. Specialist software for integrating scientific simulators with probabilistic programming has also been recently developed \citep{baydin2019etalumis} to facilitate such simulation-based inference.   

\subsection{Generative modelling and simulations}

Deep generative models, which include variational autoencoders and generative adversarial networks \textcolor{blue}{(GANs)} have been hugely successful in generating  realistic synthetic images or text similar to real-world training datasets. Increasingly simulation models are being coupled within a generative framework with very promising results. For example, molecular dynamics simulations and deep generative models are used in \citet{das2021accelerated} to accelerate antimicrobial discovery. Promising results are seen when simulation-based approaches \textcolor{blue}{are} coupled with deep generative models, as for example in inverse design of meta-materials \citep{ma2019probabilistic}, fluid simulations \citep{kim2019deep} and multi-phase flows \citep{zhong2019predicting} among others. 

\subsection{Simulation-based control}

\textcolor{blue}{With the advent of cheap and fast computation, there is an increasing trend of using complex simulation models for designing and tuning controllers.} Traditionally, control system design has involved system identification techniques for obtaining reduced-order state-space models of the physical system followed by using an efficient optimisation routine for obtaining the controller parameters. Reinforcement learning\textcolor{blue}{, on the other hand,} uses a simulation environment which can accept actions as control inputs from an agent and provide an output reward to the agent. The agent is trained to develop control policies, to maximise the cumulative reward, by running multiple simulations \citep{sutton2018reinforcement} in the training loop itself. \textcolor{blue}{Even though RL literature has existed for over three decades, large scale computation power in recent times has enabled RL based methods to achieve spectacular results.}
Deep Reinforcement Learning, \textcolor{blue}{in particular,} has gained traction in recent times due to impressive performance in learning to play Atari video games \citep{mnih2015human} and beating a human professional player at the game of Go \citep{silver2016mastering}. The environment can be any simulation including physics-based simulation engines.    

\color{blue}
\section{Translation to industrial applications}

Despite progress at the research level and material commercial benefits, industry adoption of data-centric engineering at scale is nascent: a combination of (i) organisational design and (ii) know-how barriers is impeding uptake. Data-centric engineering requires orders-of-magnitude increases in the number of commissioned simulation cases. When establishing simulation campaigns, however, communication between engineers, plant operators, and, increasingly, data scientists remains a manual and time-intensive process that does not scale. Engineering-heavy organisations must learn from the Big Tech companies who have successfully deployed operational ML at scale, and have truly moved simulation practice from batch designs to high-throughput continuous operations. The corresponding return on investment is recovered in accelerated product innovation, reduced time-to-market, and increased energy efficiency and safety. Making the leap to data-centric engineering, and ultimately digital twins, requires an organisational rethink of how simulations are coordinated. The emphasis shifts to automation and exposure of simulations as application programming interfaces (APIs) for wider consumption within and across engineering organisations. {\it We hence strongly encourage the inclusion of simulation in digital transformation strategies as a first step on the journey to engineering digital twins.}

Know-how that was once the domain of software engineers is now increasingly important for all engineers. Engineering settings increasingly rely on automation and ML, where basic knowledge of scripting and data handling techniques is essential. Just as chemical engineers are trained to navigate process flows through unit operations, engineers in the digital-twin era must navigate modes of data flow through distributed databases, API-based services and streaming systems \citep{kleppmann2017designing}. While IoT metadata curation, ontology design and alerts dashboarding frameworks are emerging \citep{cirillo2019standard}, the landscape for simulation-backed data-centric engineering is barren: with adoption blocked by systems that can be unintuitive to engineers, closed source or application-specific, such that new practitioners cannot exploit cross-cutting templates \citep{niederer2021scaling} or apply learnings from other engineering domains.

\begin{figure}[h!]
  \centering
  \includegraphics[width=14cm]{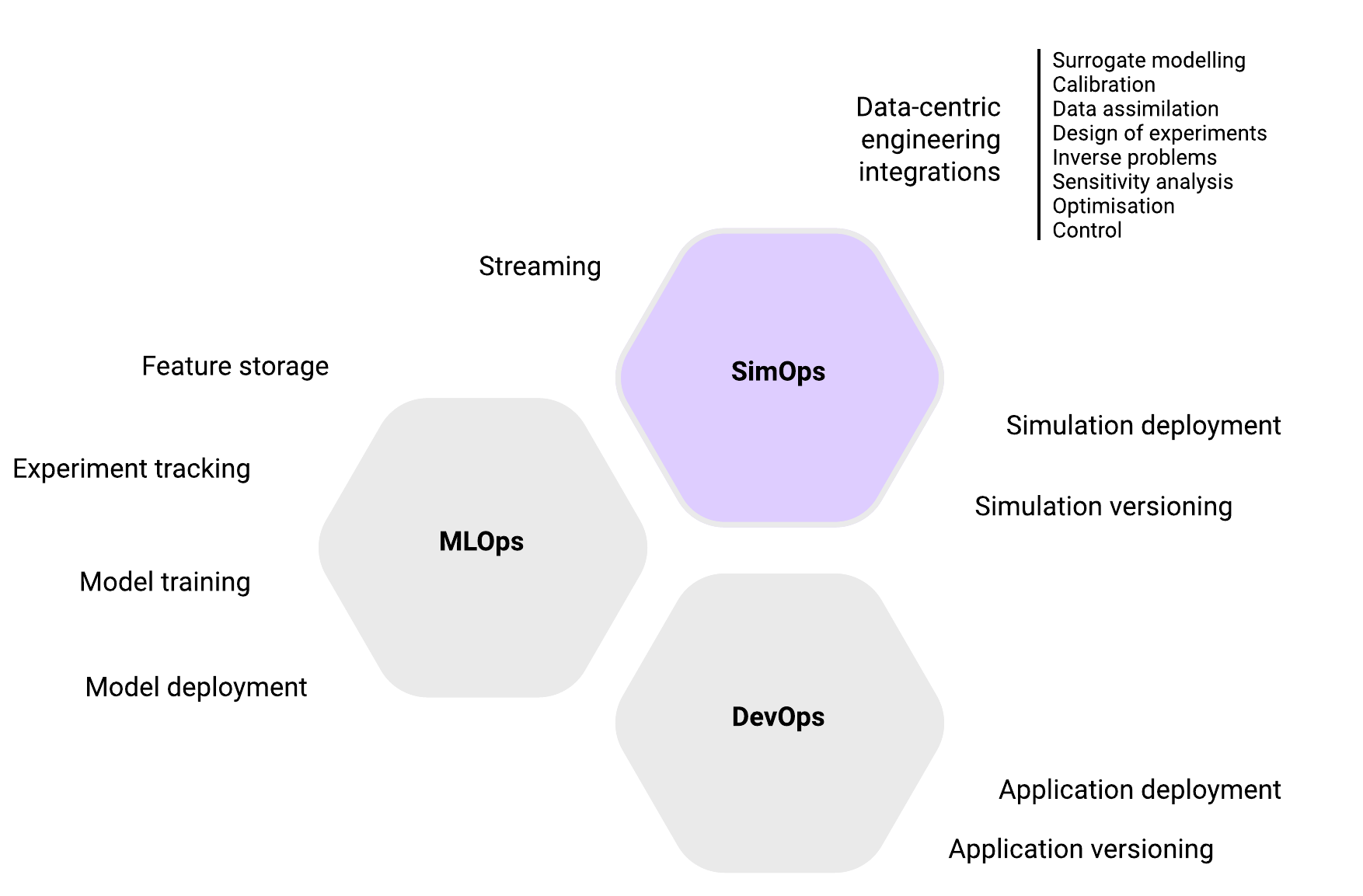}
  \caption{\textcolor{blue}{Digital twins will be powered by a confluence of SimOps, MLOps and DevOps.}}
  \label{fig:SimOps}
\end{figure}

To progress, engineering systems must learn and adopt from successful playbooks in the ML ecosystem at large. The practices of DevOps \citep{zhu2016devops}, and more recently MLOps \citep{treveil2020introducing}, are prolific in the technology industries, however, no corresponding standardised practices are in use for engineering simulations. To this end, we propose a `SimOps' framework for managing operational simulations. Digital twins will hence be powered by a confluence of all three fields: DevOps for application code, MLOps for machine learning, and SimOps for simulations lifecycle management, as illustrated in figure \ref{fig:SimOps}. As typical SimOps systems will be coupled to live real-world datastreams, engineers must be cautious of technical debt \citep{sculley2015hidden} that grows from, often deceptively simple, short-term development gains. 

We encourage the development of community-driven standards and deployment infrastructure to maximise industry uptake. Establishing a base layer of simulation automation and standardisation, through SimOps, opens the gate to high-order analyses; unlocking economic value through higher-order ecosystems of app building and digital-twin practices built on top of a robust data-centric engineering core. Organisational barriers are overcome through a multi-way democratisation of simulations; where data-science teams now have access, simulation teams unlock a data-centric engineering layer, and experimentalists are empowered by new tooling possibilities. Know-how sharing is similarly boosted through simulation versioning, historical audit trails, and collaborative editing. Typically to date, cloud based simulation has been hard-coded to specific simulation tools or locked into a specific simulation technique; SimOps hence must encompass a standard such that \textit{any} simulation tool can be linked into the higher-order ecosystem of data-centric engineering (cf.~Section~\ref{sec:integration_approaches}) in a plug-and-play fashion unlocking the vision’s full democratisation potential.

\color{black}

\section{Discussions and vision challenges}

The convergence of simulations, ML, and statistical algorithms coupled with hardware improvements including \textcolor{blue}{high power graphics processing units (GPUs)}, high computing power, cheap streaming sensors and low-cost storage is likely to have a transformative impact on traditional engineering disciplines. At the engineering design stage, the value addition will include improvements such as faster product prototyping, shorter time to market, ability to algorithmically generate and explore multiple design spaces and solutions, data and simulation driven `what if?' scenarios for effective decision making. At the engineering operations stage, improvements will include integrated simulation and data-driven solutions for better process optimisation, equipment monitoring and fault prognosis, quantitative reliability and risk assessments, operational planning and scheduling.      

There are multiple challenges that need to be overcome \textcolor{blue}{in order} to achieve a very tight integration of simulation models, statistics and machine learning. On the one hand, algorithmic advances need to be made to ensure that hybrid algorithms can leverage the best of both worlds, i.e.~retain the high predictive accuracy and computationally cheap nature of data-driven models, and at the same time incorporate elements of interpretability, encoding of physical laws and trustworthiness of simulation models. On the other hand, easy-to-use software implementations of the same should be available for wider uptake in allied fields and industrial use cases. For example, in the field of ML, deep-learning software \textcolor{blue}{such as} TensorFlow \citep{abadi2016tensorflow} and Keras \citep{chollet2016keras} has essentially democratised the technology to ensure novice users can easily adapt underlying codes and apply them to their specific use cases within a very short turnaround time.

From the industrial uptake perspective, there needs to be awareness of what impact such an integrated vision might have, the \textcolor{blue}{human-resource} requirements to execute the project, approximate project completion \textcolor{blue}{timelines}, expected outcomes, and return on investment that such a project can bring. Only then can \textcolor{blue}{such data-centric approaches} lead to effective adoption and proliferation for industrial use cases.

There needs to be upskilling among the existing industrial workforce to be able to adopt these methodologies in their daily workflows, to improve productivity, efficiency, and shorten project delivery timelines. On a longer time frame, it is important to \textcolor{blue}{refine} the university curriculum and train engineers who are data science and simulation literate from the outset. 

\section{Conclusions}

Both algorithmic and hardware improvements are paving the way towards the convergence of simulations, statistical methods and machine learning: hence we have reviewed both pre-existing and emerging use cases at this research intersection. 
Alongside existing use cases, which are becoming more powerful due to availability of computing resources, newer algorithmic  improvements are emerging that can harness the best of both modelling paradigms: simulations and data-driven learning. Though there are still multiple challenges to be overcome before realising the full potential of such technologies, democratising software solutions and upskilling industry practitioners can already have an immediate impact on engineering design and operations. Given current interest in this research area among diverse academic communities, and impactful translational opportunities for industry, such integrated approaches have enormous potential to bring about a step change in traditional engineering design and operations. \\

\section*{Acknowledgements}
\thispagestyle{empty}

We acknowledge funding from the Engineering and Physical Sciences Research Council, UK, through the Programme Grant PREMIERE (EP/T000414/1), as well as funding through the Wave 1 of The UKRI Strategic Priorities Fund under the EPSRC Grant EP/T001569/1, particularly the \emph{Digital Twins for Complex Engineering Systems} theme within that grant, and the Royal Academy of Engineering through their support of OKM's PETRONAS/RAEng Research Chair in Multiphase Fluid Dynamics.

\bibliographystyle{plainnat}
\bibliography{references}  






\end{document}